\documentclass{article}
\usepackage{spconf,amsmath,graphicx,hyperref}
\usepackage{amsmath,amsfonts,amssymb}
\usepackage{graphicx}
\usepackage{subcaption}
\usepackage{booktabs}
\usepackage{cite}
\usepackage{multirow}
\usepackage{tikz}
\usetikzlibrary{shapes.geometric,arrows.meta,positioning,calc}
\usepackage[T1]{fontenc}
\usepackage{lmodern}
\usepackage{amstext}
\usepackage{url}
\usepackage{bbm}
\usepackage{float}
\usepackage{adjustbox}
\usepackage{threeparttable}
\usepackage{array}
\usepackage{xspace}
\usepackage{pgfplots} 
\pgfplotsset{compat=1.17} 

\title{Improving Active Learning for Melody Estimation by Disentangling Uncertainties}
\name{Aayush Jaiswal$^{*}$\thanks{$^{*}$Equal contribution}, Parampreet Singh$^{*}$\footnotemark[1], Vipul Arora}
\address{Indian Institute of Technology, Kanpur}
\begin{document}
%
\maketitle
\begin{abstract}

Estimating the fundamental frequency, or melody, is a core task in Music Information Retrieval (MIR). 
Various studies have explored signal processing, machine learning, and deep-learning-based approaches, with a very recent focus on utilizing uncertainty in active learning settings for melody estimation.
However, these approaches do not investigate the relative effectiveness of different uncertainties.
In this work, we follow a framework that disentangles aleatoric and epistemic uncertainties to guide active learning for melody estimation. Trained on a source dataset, our model adapts to new domains using only a small number of labeled samples.
Experimental results demonstrate that epistemic uncertainty is more reliable for domain adaptation with reduced labeling effort as compared to aleatoric uncertainty.

\end{abstract}

\section{Introduction}
\label{sec:intro}

Estimating the fundamental frequency, or melody, is a central problem in Music Information Retrieval (MIR). 
It underpins several downstream applications, including music search, transcription, generation, and recommendation \cite{salamon2012melody, bittner2017deep}. 
Early methods for melody estimation relied on signal-processing heuristics \cite{salamon2012melody,mauch2014pyin}, followed by
deep learning approaches
framing it as a classification task over discretized pitch bins \cite{bittner2017deep,lu2018automatic,hsieh2019streamlined,conf_interspeech_GaoHWHH23,yu2021frequency}, or as a regression task
\cite{saxena2025regression,jing2025joint}.
Recent studies \cite{saxena2024interactive,saxena2025regression} have explored uncertainty estimation for melody estimation in an active learning setting.
However, they rely on aggregate uncertainty, without disentangling it into various factors.

Model uncertainties can be factorized into two factors \cite{kendall2017uncertainties}: \emph{aleatoric uncertainty}, which arises from inherent data ambiguity and cannot be reduced, and \textit{epistemic uncertainty}, which reflects model uncertainty and can be reduced by collecting additional informative data samples. 
Numerous works have explored uncertainty estimation in deep models \cite{gal2016dropout,lakshminarayanan2017simple,corbiere2019addressing_TCP,sumit_param_confidence_TCP,kendall2017uncertainties,amini2020deep,sensoy2018evidential,maddox2019simple,Wang2018AleatoricUE,depeweg2018decomposition_infotheory}. 
Monte Carlo dropout \cite{gal2016dropout} estimates epistemic uncertainty through multiple stochastic forward passes, with dropout layer inducing different model realizations on each pass.
\cite{Wang2018AleatoricUE} extends this framework to also capture aleatoric uncertainty. Deep ensembles \cite{lakshminarayanan2017simple} model epistemic uncertainty through disagreement among independently trained networks, with each network additionally predicting aleatoric variance. Bayesian neural networks provide joint estimates of both types of uncertainty \cite{ kendall2017uncertainties,depeweg2018decomposition_infotheory}, but at the cost of significant computational overhead. In contrast, evidential deep learning \cite{amini2020deep,sensoy2018evidential} learns a higher-order evidential distribution that allows both aleatoric and epistemic uncertainties to be obtained in a single forward pass. It is computationally efficient and avoids the need for ensembles or out-of-distribution data.

In this work, we investigate if disentangling uncertainty can help improve melody estimation in an active learning setting. 
We train a higher-order distribution that estimates aleatoric and epistemic uncertainties for melody estimation under both regression and classification settings. Further, we use these uncertainties for active learning. Experiments demonstrate that epistemic uncertainty is more effective for active data selection than aleatoric uncertainty, which makes it very useful for low-resource melody estimation. 
Our code is available at:
\url{https://github.com/AayushJaiswal01/melody-extraction-evidential}

\section{Preliminaries}

Consider an input-label pair $(x,y)$, where $y$ denotes the frame-wise melody label for the input audio $x$. 
A model is trained to predict both the melody as well as aleatoric and epistemic uncertainties separately. This is done by placing a higher-order prior over the likelihood parameters that computes the mean and uncertainty components in a single forward pass \cite{sensoy2018evidential,amini2020deep}.

In the classification setting, the label $y \in \{1,...,K\}$ is categorical in nature. We assume class labels are drawn from a Categorical distribution, $y \sim \textit{Categorical}(\mathbf{p})$, whose probabilities $\mathbf{p}$ are modeled with a Dirichlet prior, $\mathbf{p} \sim \text{Dir}(\mathbf{p}|\boldsymbol{\alpha})$ \cite{sensoy2018evidential}. The model outputs the evidence vector $\boldsymbol{\alpha}$, with total evidence given by $S = \sum_k \alpha_k$. The estimated melody corresponds to the class with the highest mean probability, $\hat{p}_k = \alpha_k / S$. Aleatoric uncertainty ($u_a$) and epistemic uncertainty ($u_e$) are obtained from the entropy decomposition of the Dirichlet distribution:
\[
u_a = \sum_{k=1}^K \frac{\alpha_k}{S} \big( \psi(S+1) - \psi(\alpha_k+1) \big), 
\]
\[
u_e = -\sum_{k=1}^K \frac{\alpha_k}{S}\log\left(\frac{\alpha_k}{S}\right) - u_a
\]
where $\psi(\cdot)$ is the digamma function.

In case of regression, the target $y \in \mathbb{R}$ is assumed to follow a Gaussian likelihood \cite{amini2020deep}. The model assumes:
\[
y \sim \mathcal{N}(\mu, \sigma^2), \quad (\mu, \sigma^2) \sim \text{NIG}(\gamma, \nu, \alpha, \beta),
\]  
where $\gamma \in \mathbb{R}$, $\nu > 0$, $\alpha > 1$, $\beta > 0$, and NIG represents the Normal-Inverse Gamma distribution. The parameter $\gamma$ represents the estimated melody, and aleatoric uncertainty ($\sigma_a^2$) and epistemic uncertainty ($\sigma_e^2$) are given by \cite{amini2020deep}: 
\[
\sigma_a^2 = \frac{\beta}{\alpha - 1}, 
\quad 
\sigma_e^2 = \frac{\beta}{\nu(\alpha - 1)}.
\]  
In this work, we adopt similar training settings for both classification and regression tasks and evaluate them alongside established baselines for active learning.

\section{Method}\label{Sec:Methodology}

\subsection{Problem Formulation}

Given a mel-spectrogram $\mathbf{X} \in \mathbb{R}^{T \times F}$, we formulate joint voicing detection and fundamental frequency estimation $f_0(t)$. The pitch range [51.91, 830.61] Hz is discretized into $K = 384$ logarithmic bins with 12.5 cents resolution following the approach in \cite{saxena2025regression}. A binary cross-entropy head performs voiced/unvoiced classification for all frames. For unvoiced frames, training relies solely on the voicing classification loss. For voiced frames, the model employs a dual formulation where the same bin centers serve different purposes. In the classification approach, these bin centers constitute $K$ categorical target classes for Dirichlet parameterization through evidence scores $\boldsymbol{\alpha}$, while in the regression approach, they serve as continuous targets for training NIG parameters.

\subsection{Model Architecture}
A ResNet model with four convolutional blocks with filter sizes (32, 64, 128, 256) using bottleneck layers, batch normalization, LeakyReLU, residual connections, and max-pooling. Dropout (0.3) and L2 regularization ($10^{-5}$) are applied for regularization. The same model is used for all methods, with output heads varying based on the problem setting.

\begin{table*}[!tb] 
\centering
\begin{threeparttable}
\caption{Cross-dataset performance comparison. Values show RPA/RCA/OA (\%). FT indicates fine-tuning with samples selected based on epistemic uncertainty (for M1 and M2) and TCP confidence scores for TCP (FT).}
\label{tab:main_results}
\small
\setlength{\tabcolsep}{4.5pt} 
\renewcommand{\arraystretch}{1.1}
\begin{tabular}{|l|ccc|ccc|ccc|ccc|}
\hline
\multirow{2}{*}{\textbf{Method}} & \multicolumn{3}{c|}{\textbf{MIR-1K}} & \multicolumn{3}{c|}{\textbf{HAR}} & \multicolumn{3}{c|}{\textbf{ADC2004}} & \multicolumn{3}{c|}{\textbf{MIREX-05}} \\
\cline{2-13}
& \textbf{RPA} & \textbf{RCA} & \textbf{OA} & \textbf{RPA} & \textbf{RCA} & \textbf{OA} & \textbf{RPA} & \textbf{RCA} & \textbf{OA} & \textbf{RPA} & \textbf{RCA} & \textbf{OA} \\
\hline
$\beta$-NLL (Base) & 71.8 & 72.4 & 53.3 & 66.1 & 66.4 & 58.2 & 42.6 & 45.0 & 36.3 & 74.7 & 75.5 & 60.6 \\
\hline
TCP (Base) & 81.1 & 82.3 & 84.6 & 71.0 & 71.9 & 73.1 & 43.1 & 46.2 & 46.9 & 77.4 & 78.4 & 82.0 \\
TCP (FT) & 81.2 & 82.6 & 84.4 & 81.1 & 84.8 & 83.0 & 55.3 & 59.8 & 55.2 & 79.9 & 80.4 & 83.9 \\
\hline
M1 (Base) & 75.8 & 78.5 & 81.7 & 
69.7 & 72.4 & 72.8 & 
43.7 & 47.2 & 47.9 & 
71.8 & 74.0 & 78.9 \\

M1 (FT) & 76.1 & 78.5 & 80.7 & 85.7 & 88.1 & 86.9 & 
59.0 & 68.8 & 52.5 & 
78.9 & 81.1 & 81.5 \\
\hline
\textbf{M2 (Base)} & 80.9 & 81.3 & 84.6 & 66.8 & 67.7 & 69.2 & 44.0 & 46.0 & 47.1 & 78.3 & 79.2 & 82.5 \\
\textbf{M2 (FT)} & 81.9 & 82.6 & 85.3 & \textbf{96.2} & \textbf{96.3} & \textbf{96.0} & \textbf{68.8} & \textbf{70.0} & \textbf{64.4} & \textbf{85.0} & \textbf{85.4} & \textbf{87.1} \\
\hline
\end{tabular}
\begin{tablenotes}
\item All base models were trained on MIR-1K and tested as it is on other datasets. Fine Tuning (FT) uses N=1000 samples for MIR-1K/HAR and N=100 samples for ADC2004/MIREX-05.
\end{tablenotes}
\end{threeparttable}
\end{table*}

\subsection{Classification Objective (M1)}

For classification, the loss $L_{\text{M1}}$ follows the Type-II Maximum Likelihood objective \cite{sensoy2018evidential}:
\[
L_{\text{M1}} = \frac{1}{\sum_i v_i} \sum_{i} v_i \cdot \left( L_{\text{NLL}}(\boldsymbol{\alpha}_i, \mathbf{y}_i) + \lambda_t \, L_{\text{KL}}(\boldsymbol{\alpha}_i, \mathbf{y}_i) \right)
\]
Here, 
$$L_{\text{NLL}} = \sum_k y_{ik} ( \psi(S_i) - \psi(\alpha_{ik}) )$$ is the negative log-likelihood, which maximizes evidence $\boldsymbol{\alpha}_i$ for the correct class $\mathbf{y}_i$ and  $v_i \in \{0,1\}$ indicates whether frame $i$ is voiced (1 if voiced, 0 if unvoiced). Where $S_i = \sum_k \alpha_{ik}$ and $\psi(\cdot)$ is the digamma function. The term $L_{\text{KL}}$ is a regularizer defined by the KL-divergence between the model's posterior and a uniform Dirichlet distribution, penalizing spurious evidence for incorrect classes \cite{sensoy2018evidential}. The coefficient $\lambda_t$ is annealed during training. At inference, frames predicted as unvoiced are set to $f_0=0$, while for voiced frames, the pitch is the class with the highest mean probability.
Total loss is given by:
\begin{equation}
\label{eq:total_loss}
L_{\text{c}} = L_{\text{BCE}} + w \cdot L_{\text{M1}}
\end{equation}
where $L_{\text{BCE}}$ is the BCE loss for classification of voiced vs unvoiced frames.
\subsection{Regression Objective (M2)}
For regression (M2), the loss $L_{\text{M2}}$ is the evidential regression loss applied to voiced frames ($v_i=1$):
\[
L_{\text{M2}} = \frac{1}{\sum_i v_i} \sum_i v_i \cdot \Big( L_{\text{NLL},i} + \lambda L_{\text{R},i} \Big)
\]
Here, $L_{\text{NLL},i}$ enforces data fidelity by maximizing the likelihood of the ground-truth frequency $y_i$ under the predicted Normal-Inverse Gamma (NIG) distribution. The regularizer $L_{\text{R},i}$ discourages confident predictions for large errors and is defined as \cite{amini2020deep}:
\[
L_{\text{R},i} = | y_i - \gamma_i | (2\nu_i + \alpha_i)
\]
where $\gamma_i$ , $\nu_i$ and $\alpha_i $ are parameters of the predicted NIG distribution.
The total loss is given by:
\begin{equation}
\label{eq:total_loss}
L_{\text{reg}} = L_{\text{BCE}} + w \cdot L_{\text{M2}}
\end{equation}

At inference, frames predicted unvoiced are set to $f_0=0$, while voiced frames use $\hat{f}_0 = \gamma_i$ with associated aleatoric and epistemic uncertainties.
\subsection{Active Learning}
We use uncertainty-driven active learning to adapt models across domains. For each sample, frame-level uncertainty values are computed and averaged across all frames to obtain a sample-level uncertainty score. The top-$K$ most uncertain samples are then selected for fine-tuning (FT).

\subsection{Evaluation Metrics}
Performance is assessed using standard metrics via the \texttt{mir\_eval} library \cite{raffel2014mir_eval}: Overall Accuracy (OA), measuring the fraction of correctly estimated melody frames; Raw Pitch Accuracy (RPA), capturing frames with exact pitch match; and Raw Chroma Accuracy (RCA), capturing pitch matches modulo octave.

\section{Experimental Results}

\subsection{Experimental Setup}
All audio is uniformly preprocessed by converting to mono, downsampling to 16 kHz, and segmenting into non-overlapping 1-second clips. For each audio clip, a log-magnitude spectrogram, computed with a 2048-point STFT and a 10 ms hop size, serves as the model input.

\subsubsection{Datasets}
MIR-1K\footnote{[Online]. Available: https://sites.google.com/site/unvoiced
soundseparation/mir-1k} dataset consisting of 1000 Chinese karaoke excerpts with clean vocals mixed with accompaniments, totaling approximately 2.2 hours of audio forms the source domain for our task. 
For FT, we consider three diverse target domains. The Hindustani Alankaar and Raga (HAR) dataset \cite{saxena2024interactive}, containing 523 audio recordings of Indian classical singing with a total duration of 6.84 hours. We partition the dataset such that the training set contains recordings from one professional singer, while the test set consists of recordings from the other. ADC2004\footnote{\label{footnote_1}[Online]. Available: http://labrosa.ee.columbia.edu/projects/melody/}, comprises 12 excerpts of Western pop music, while MIREX-05\footref{footnote_1} includes 9 multi-genre excerpts.
These two are relatively small in size compared to MIR-1k and HAR, yet they remain standard benchmarks in melody estimation.
For the source domain, we adopt a 70/15/15 train/validation/test split, whereas for all target domains, we use an 80/20 train/test split.

\subsubsection{Baselines}
\textbf{$\beta$-NLL Regression}
The first baseline follows the heteroscedastic regression formulation. The network outputs a predicted mean $\hat{\mu}_i$ and log variance $\hat{\sigma}_i^2$ for each input frame. 
The model is trained with the $\beta$-NLL objective:
\begin{equation}
L_{\beta\text{-NLL}} = \frac{1}{N} \sum_{i} (\hat{\sigma}_i^2)^{\beta} 
\left( \frac{(y_i - \hat{\mu}_i)^2}{2\hat{\sigma}_i^2} + \tfrac{1}{2}\log(\hat{\sigma}_i^2) \right).
\end{equation}
Uncertainty is derived from the predicted variance $\hat{\sigma}_i^2$, which conflates aleatoric and epistemic components. 

\textbf{TCP Confidence Scores (classification)}
The second baseline adopts the modified True Class Probability (TCP) framework~\cite{saxena2024interactive}. The pitch range (51.91–830.61 Hz) is discretized into 384 logarithmic bins (12.5 cents per bin), and a ResNet classifier is trained over these bins using categorical cross-entropy. An auxiliary confidence head is then trained to predict normalized TCP values with mean-squared error loss, keeping the model parameters frozen. 

\subsection{Results}

Table~\ref{tab:main_results} presents the results across source and target domains. All models experience a drop in their performance in the new target domain, showing the difficulty of domain shift.
For classification, the TCP method shows comparable base and FT results compared to M1, even surpassing M1 in some cases. But because the uncertainties are still entangled, it does not give such large performance boost for FT as is seen in M2.
We find that M1 and M2 achieve similar base performance with quantized targets, but M2 yields stronger FT gains when guided by epistemic uncertainty, indicating that regression provides a clearer disentanglement of uncertainties.
It also highlights that epistemic uncertainty is the most reliable signal for sample selection while aleatoric uncertainty contributes a little towards FT.  

\subsubsection{Ablation Studies}
For model training, we carry out ablation studies comparing three configurations to get the best model for regression. R1 is purely regression, treating the target values as their actual frequency values, without quantizing. R2 is regression on quantized 384-bin targets without explicit voicing separation, and our final model M2 with voicing detection, along with bin quantization, which has already been discussed in Section~\ref{Sec:Methodology}. We carry out ablation studies on base model itself, to find out the best model for regression, which then undergoes FT.
As we can see in Table~\ref{tab:ablation}, R1 performs poorly, while R2 improves for MIR-1K. Both show similar results for HAR. 
As expected, M2 gives best results and therefore we adopt M2 for FT.

\subsubsection{Fine-Tuning with Active Learning}
\begin{figure}[t!]
\centering
\begin{tikzpicture}
\begin{axis}[
    width=0.97\columnwidth, 
    height=6cm,
    xlabel={Number of Fine-tuning samples (N)},
    ylabel={Overall Accuracy (\%)},
    xmin=0, xmax=1000,
    ymin=35, ymax=100,
    xtick={0, 200, 400, 600, 800, 1000},
    ytick={40, 50, 60, 70, 80, 90, 100},
    ymajorgrids=true,
    grid style=dashed,
    line width=0.6pt,
    legend style={
        at={(0.5,-0.3)},    
        anchor=north,        
        legend columns=3,    
        font=\footnotesize,
        cells={anchor=west}
    }
]
\definecolor{colorEpistemic}{RGB}{28,120,181} 
\definecolor{colorConfidence}{RGB}{214,39,40}  
\definecolor{colorAleatoric}{RGB}{255,127,14} 
\definecolor{colorVariance}{RGB}{44,160,44}   
\definecolor{colorPurple}{RGB}{148,103,189} 
\definecolor{colorCyan}{RGB}{23,190,207} 

\addplot[color=colorVariance]
    coordinates { (0, 58.2)(100, 49.4)(200, 50.4)(300, 54.1)(400, 57.3)(500, 59.3)(600, 64.1)(700, 65.3)(800, 66.7)(900, 66.9)(1000, 70.2) };
    \addlegendentry{$\beta-NLL$}

\addplot[color=colorConfidence]
    coordinates { (0, 73.1)(100, 80.6)(200, 83.6)(300, 81.4)(400, 82.4)(500, 83.5)(600, 81.5)(700, 81.9)(800, 83.8)(900, 83.5)(1000, 83.0) };
    \addlegendentry{TCP}
    
\addplot[color=colorPurple]
    coordinates { (0, 72.84)(100, 67.98)(200, 76.01)(300, 78.99)(400, 82.24)(500, 82.72)(600, 83.02)(700, 84.54)(800, 85.03)(900, 85.84)(1000, 85.83) };
    \addlegendentry{M1(A)}

\addplot[color=colorCyan]
    coordinates { (0, 72.84)(100, 80.43)(200, 80.69)(300, 83.72)(400, 83.28)(500, 84.58)(600, 85.51)(700, 85.75)(800, 86.55)(900, 86.75)(1000, 86.94) };
    \addlegendentry{M1 (E)}

\addplot[color=colorAleatoric]
    coordinates { (0, 69.2)(100, 78.1)(200, 78.8)(300, 75.5)(400, 75.5)(500, 73.2)(600, 71.2)(700, 70.4)(800, 69.9)(900, 69.1)(1000, 71.6) };
    \addlegendentry{M2 (A)}

\addplot[color=colorEpistemic]
    coordinates { (0, 69.2)(100, 91.2)(200, 93.6)(300, 94.6)(400, 94.9)(500, 95.3)(600, 95.5)(700, 95.5)(800, 95.5)(900, 95.8)(1000, 96.0) };
    \addlegendentry{M2 (E)}
 
\end{axis}
\end{tikzpicture}
\caption{Overall Accuracy on the HAR dataset vs the number of samples used for fine-tuning. (A) represents Aleatoric and (E) represents Epistemic Uncertainty}
\label{fig:har_al_comparison}
\end{figure}
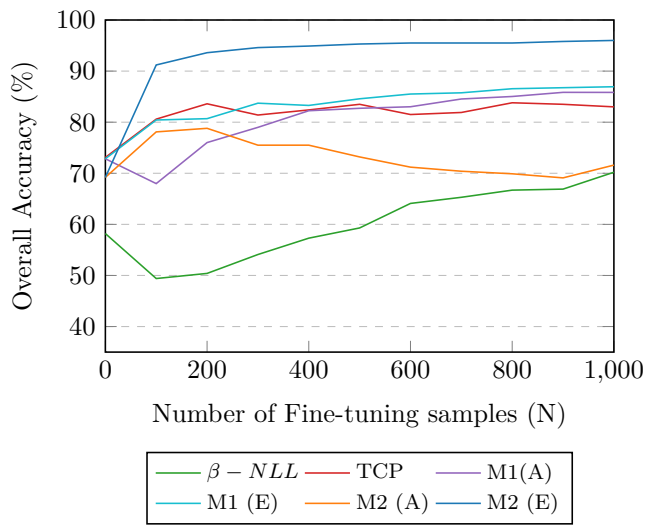

We study how different types of uncertainties help in active learning by gradually increasing the number of target-domain samples, selecting $K \in \{100, 200, \dots, 1000\}$ most uncertain samples from HAR for FT. Figure~\ref{fig:har_al_comparison} shows the adaptation curves. For M1, both aleatoric and epistemic uncertainties lead to similar improvements. This can be because of the fact that for classification, the uncertainties are not well disentangled and remain somewhat correlated. In contrast, for M2, epistemic uncertainty provides a clear advantage, yielding over 10\% higher accuracy than both M1 and M2's own aleatoric uncertainty variant. This demonstrates the benefit of formulating melody estimation as a regression task and the effectiveness of epistemic uncertainty for active learning, as it is able to achieve promising performance using just 200 samples out of the target domain.

\begin{table}[!tb] 
\centering
\begin{threeparttable}
\caption{Ablation study on MIR-1K (source) and HAR (target) base models, without Fine Tuning}
\label{tab:ablation}
\small
\setlength{\tabcolsep}{4.5pt} 
\renewcommand{\arraystretch}{1.1}
\begin{tabular}{|l|ccc|ccc|}
\hline
\multirow{2}{*}{\textbf{Method}} & \multicolumn{3}{c|}{\textbf{MIR-1K}} & \multicolumn{3}{c|}{\textbf{HAR}} \\
\cline{2-7}
& \textbf{RPA} & \textbf{RCA} & \textbf{OA} & \textbf{RPA} & \textbf{RCA} & \textbf{OA} \\
\hline
R1 & 56.0 & 56.5 & 66.7 & 46.0 & 46.1 & 51.1 \\
R2 & 70.9 & 71.7 & 76.2 & 47.2 & 49.0 & 50.6 \\
\textbf{M2} & \textbf{80.9} & \textbf{81.3} & \textbf{84.6} & \textbf{66.8} & \textbf{67.7} & \textbf{69.2} \\
\hline
\end{tabular}
\end{threeparttable}
\end{table}

\section{Conclusion}

In this work, we introduce an uncertainty-guided framework for active melody estimation. Our approach models regression and classification for melody estimation along with voicing detection, showing promising gains.
We disentangle epistemic and aleatoric uncertainties and test their usefulness in an active learning setting.
Experiments highlight that epistemic uncertainty serves as a far better approach for active learning.
A model trained on a source domain (data rich) can be adapted to unseen target domains (data poor) using only a small number of labeled target samples selected based on epistemic uncertainty. This achieves strong performance while reducing labeling and computational costs. 
These findings underscore the advantages of uncertainty disentanglement and the central role of epistemic uncertainty in cross-domain melody estimation.

\begingroup
\small  
\bibliographystyle{IEEEbib}
\bibliography{refs}
\endgroup
\end{document}